\documentclass[twocolumn,showpacs,preprintnumbers,prd,nofootinbib]{revtex4-2}
\usepackage{feynman}
\usepackage{slashed}
\usepackage{graphicx,amsmath,amssymb,xcolor,hyperref}

\begin{document}

\author{Yi Chung}
\email{yi.chung@mpi-hd.mpg.de}
\affiliation{Max-Planck-Institut f\"ur Kernphysik, Saupfercheckweg 1, 69117 Heidelberg, Germany}

\title{Naturalness-motivated composite Higgs
model for generating the top Yukawa coupling}

\begin{abstract}
The large top Yukawa coupling results in the top quark contributing significantly to the quantum correction of the Higgs mass term. Traditionally, this effect is canceled by the presence of top partners in symmetry-based models. However, the absence of light top partners poses a challenge to the Naturalness of these models. In this paper, we study a model based on composite Higgs with the top Yukawa coupling originating from dimension-six four-fermion operators. The low cutoff scale of the top quark loop required by the Naturalness principle can be realized with a light gauge boson $E_\mu$ which connects the hyperfermions and top quarks. A scalar-less dynamical model with weakly coupled extended $SU(4)_{EC}$ gauge group is presented. The model features an $E_\mu$ boson and a $Z'_E$ boson both at the sub-TeV scale, which lead to a rich phenomenology, especially in the top physics.
\end{abstract}

\maketitle

\section{Introduction}

The Standard Model (SM) of particle physics successfully describes all known elementary particles and interactions. At the center of SM is the mechanism of electroweak symmetry breaking (EWSB), which is responsible for the masses of SM gauge bosons and fermions. The discovery of Higgs bosons in 2012 \cite{Chatrchyan:2012xdj, Aad:2012tfa} filled in the last missing puzzle of the SM. Nevertheless, the SM does not address the UV sensitivity of the Higgs boson mass, which is known as the hierarchy problem. The Higgs mass term receives divergent radiative corrections from the interactions with SM fields, especially the top quark due to its large Yukawa coupling. The contribution can be derived numerically by calculating the one-loop diagram with the top quark and is given by
\begin{align}\label{toploop0}
\Delta m_H^2|_{\text{top}}
&\sim-i\,2N_c\,y_t^2 \int_{}^{} \frac{d^4k}{(2\pi)^4}\frac{k^2+m_t^2}{(k^2-m_t^2)^2}\nonumber\\
&= -\frac{3}{8\pi^2}y_t^2
\left[\Lambda_{t}^2-3\,m_t^2\,\text{ln}\left(\frac{\Lambda^2_{t}}{m_t^2}\right)+\cdots\right]~,
\end{align}
where $\Lambda_{t}$ is the scale of the top Yukawa coupling.

To avoid the large quadratic corrections, most models invoke new symmetry such that the corrections cancel in the symmetric limit. New degrees of freedom, known as top partners, are introduced to cancel out the $\Lambda^2_{t}$ term. However, the symmetry can not be exact and the difference between the top mass $m_t$ and top partner mass $M_T$ will reintroduce the correction given by
\begin{align}\label{toploop+}
\Delta m_H^2|_{\text{top}}+\Delta m_H^2|_{\text{top partner}}
\sim -\frac{3}{8\pi^2}y_t^2 M_T^2~.
\end{align}
Following the Naturalness principle \cite{Giudice:2008bi,Giudice:2013yca,Craig:2022eqo}, we expect top partners to show up at the sub-TeV scale to avoid fine-tuning. However, after years of searches, the bounds of colored top partner mass $M_T$ have reached $1.5$ TeV for both scalar partners \cite{CMS:2020pyk, ATLAS:2020aci} and fermionic partners \cite{CMS:2019eqb, CMS:2022fck, ATLAS:2018ziw, ATLAS:2022ozf, ATLAS:2022hnn}. The non-observation of colored top partners thus poses a challenge to the naturalness of these types of models.

In this study, we focus on an alternative scenario \cite{Bally:2023lji} where the top Yukawa coupling originated from dimension-six operators with a scale $\Lambda_t$. If we can have the scale $\Lambda_t \lesssim 1$ TeV, the contribution from the top loop will be under control. The idea has already been realized at the one-loop level in \cite{Bally:2022naz} with an elementary Higgs and top quark. In this paper, instead, we consider that the observed Higgs boson is a composite state \cite{Kaplan:1983fs, Kaplan:1983sm} formed by hyperfermions from a strongly coupled theory.

Generating SM Yukawa couplings in a strongly coupled theory can be traced back to Extended Technicolor (ETC) \cite{Dimopoulos:1979es, Eichten:1979ah, Farhi:1979zx}, where SM Yukawa couplings arise from dimension-six four-fermion operators. The scale $\Lambda_t$ is determined by the mass of new massive gauge bosons $\Lambda_t\sim M_{ETC}$ which connect the hyperfermions and SM fermions. The models based on modern composite Higgs models have also been studied in \cite{Cacciapaglia:2015yra}. However, for the generic mass $M_{ETC} \sim g_{E}f_{E}$, the breaking scale $f_{E}$ is fixed by the value of the top Yukawa coupling at around the TeV scale and $g_{E}$ is the coupling of the ETC group which is related to the strong coupling responsible for the hyperfermion condensate so the mass $M_{ETC}$ is expected to be heavy from the theoretical aspect.

Motivated by the Naturalness principle, we aim at a model with a small $g_E$ such that the scale $\Lambda_t$ can be low. That is, the gauge group that connects hyperfermions and top quarks is weakly coupled and independent of the strong interaction. Moreover, we want to construct a fully dynamical model, where the two relevant scales, $f$ and $f_E$, both come from strong dynamics. We will show how to get all these features in a fermionic theory with an extended gauge group. The phenomenology is also presented with a special focus on the top physics.

This paper is organized as follows. In Sec.~\ref{sec:Basic}, we introduce the basic idea and issue in an ETC-like mechanism and how we are going to solve them. Starting with the extension of the gauge group in Sec.~\ref{sec:Extend}, we briefly go through the difference between the traditional way and the new way we work on. A concrete model is presented in Sec.~\ref{sec:Model} with three relevant mechanisms discussed in detail. The important phenomenology is presented, including the indirect searches in Sec.~\ref{sec:Indirect} and direct searches in Sec.~\ref{sec:Direct}. Sec.~\ref{sec:Conclusion} contains conclusions and outlooks.

%\newpage

\section{Basic idea and issue of top Yukawa from Four-Fermion Operators}\label{sec:Basic}

To generate the top Yukawa from dimension-six four-fermion operators, we need to first introduce an extended gauge group $\mathcal{G}_{E}$ with gauge bosons $G_E^a$ and coupling $g_E$, where the top quarks and hyperfermions $\psi$ are within the same multiplets $Q$. The generic Lagrangian is given by 
\begin{align}
{\cal L}_{\text{E}}=& \,g_{E} G_{E,\mu}^{a} 
(\bar{Q}_L\gamma^\mu T^a Q_L+\bar{Q}_R\gamma^\mu T^a Q_R)\nonumber\\
\supset& \,\frac{1}{\sqrt{2}} g_{E} {E}_{\mu} (\bar{\psi}_L\gamma^\mu q_L+\bar{\psi}_R\gamma^\mu t_R)\,,
\end{align}
where $E_\mu$ is the specific boson among $G_E^a$ that mediates the top quarks and hyperfermions. The group $\mathcal{G}_{E}$ is then broken at the scale $f_{E}$ down to the SM gauge group $\mathcal{G}_{SM}$ and hypercolor  $\mathcal{G}_{HC}$ (can be either broken or unbroken)\footnote{In this study, we use the term hypercolor instead of technicolor to refer to the strong interaction, as is commonly used in modern composite Higgs models. In addition to the conventional confining hypercolor, we also consider the scenario where hypercolor is broken, resulting in a nonconfining strong interaction.}. After integrating out the massive $E_\mu$ gauge bosons with a mass $M_E$, we get an low energy effective Lagrangian as
\begin{align}
\mathcal{L}_{\text{eff}}&=-\frac{g_{E}^2}{2M_{E}^2}
\left(\bar{q}_{L}\gamma^\mu {\psi}_{L}\right)
\left(\bar{\psi}_{R}\gamma_\mu {t}_{R}\right) + h.c. \nonumber\\
&\to\frac{g_{E}^2}{M_{E}^2}
\left({\bar{\psi}_R}{\psi}_{L}\right)
\left({\bar{q}_L}{t}_{R}\right)+\cdots \text{(after Fierzing)}~.
\end{align}
Once hypercolor becomes strongly coupled and condenses the hyperfermions with a breaking scale $f$, the ${\bar{\psi}_R}{\psi}_{L}$ will form a bound state that behaves like the SM Higgs. The top Yukawa coupling is then generated with a value
\begin{align}
y_t\sim \frac{1}{v}\frac{g_{E}^2}{M_{E}^2}\langle \bar{{\psi}}_R{\psi}_{L}\rangle_{HC}
\sim \frac{g_{E}^2}{M_{E}^2}\cdot Y_s f_{}^2
\end{align}
where the coupling $Y_s$ is the Yukawa coupling from the strong dynamics with an $O(1)$ value. As the $\mathcal{G}_{E}$ is broken by $f_E$, we expect the relation $M_{E}\sim g_{E}f_{E}$ and thus
\begin{align}
y_t\sim  \left(\frac{f_{}}{f_{E}}\right)^2 Y_s \sim 1~,
\end{align}
which fixes the ratio among scales as $f_{E}\sim O(1) \times f_{}$.

Now we have a rough description for the top Yukawa coupling generated from four-fermion interactions in the composite Higgs model. However, to attain a concrete model, several issues must be addressed.

The first issue is the gauge group $\mathcal{G}_E$, which requires an extension of the SM gauge group to combine hyperfermions and top quarks into the same representation. Moreover, motivated by the Naturalness principle, we want to have a light mediator $E_\mu$. Its mass $M_E$ is given by the product of coupling $g_E$ and the breaking scale $f_{E}$. As the scale $f_{E}$ is fixed by the value of the top Yukawa coupling, we aim at a model with a small $g_E$. That is, the gauge group that connects hyperfermions and top quarks should be weakly coupled, which will be further discussed in the next section.

Second, since we aim at a fully dynamical model, the two relevant scales, $f$ and $f_E$, should both come from strong dynamics. The difference between the two scales is the key to explaining the value of top Yukawa coupling. If $f=f_E$, then $y_t\sim Y_s$ which will predict a much heavier top quark as in top condensation models \cite{Miransky:1988xi, Miransky:1989ds, Marciano:1989xd, Bardeen:1989ds, Hill:1991at}. A viable mechanism to generate a sequence of scales in a strongly coupled theory is the tumbling mechanism \cite{Raby:1979my}, which will be applied in our concrete model.

The other concern about the ETC-type models is the flavor constraints. However, given that our primary motivation is Naturalness and our goal is to lower the top loop cutoff, we assume that this mechanism is specific for top quarks and ignore the light fermions at this stage. Then, the main constraints in flavor physics will come from $B$ meson physics due to the $b_L$ inside $q_L$, which will be discussed in Sec.~\ref{sec:Indirect}.

%\newpage

\section{Extend the gauge group}\label{sec:Extend}

With the SM gauge group $SU(3)_C\times SU(2)_W\times U(1)_Y$, there are many different ways to extend it to include hyperfermions $\psi$. In this work, we focus on the cases with extended $SU(3)_C$. Other cases like extended $SU(2)_W$ are also possible and have been studied in ETC models \cite{Chivukula:1994mn} but we will leave them for future study.

\subsection{Traditional extension: $ \mathcal{G}_{HC} \times \mathcal{G}_{SM} \subset \mathcal{G}_E$}\label{sec:Traditional}

Traditional ways following the ETC models usually have the hypercolor group combined with one of the SM gauge groups to a larger group. From the top down, the extended group $\mathcal{G}_E$ group is broken down to $\mathcal{G}_{HC}\times \mathcal{G}_{SM}$ at the scale $f_E$, which separates the fermion $Q$ to the hyperfermions and top quarks.

Following the idea in \cite{Cacciapaglia:2015yra}, the hypercolor group $\mathcal{G}_{HC}=SU(N)_{HC}$  is combined with $SU(3)_C \subset \mathcal{G}_{SM}$ to $\mathcal{G}_E = SU(N+3)_{E}$. The desired fermion content $Q_{L,R}$ under $ SU(N+3)_{E}\times SU(2)_W$ is given by (we ignore the $U(1)$ in this section for simplicity)
\begin{align}
Q_L=(N+3,2),\quad Q_R=(N+3,1)~.
\end{align}
Then, the $\mathcal{G}_E$ group is broken down as 
\begin{align}
SU(N+3)_{E}\to SU(N)_{HC}\times SU(3)_C
\end{align}
After breaking, The fermions are also separated to (under $SU(N)_{HC}\times SU(3)_{C}\times SU(2)_W$)
\begin{align}
\psi_L&=(N,1,2),\quad \psi_R=(N,1,1)~\nonumber\\
q_L&=(1,3,2),~~\quad t_R=(1,3,1)~.
\end{align}
The gauge boson ${E}_\mu$, which mediates hyperfermions and top quarks, has a quantum number
\begin{align}
{E}_\mu=(N,\bar{3},1)~,
\end{align}
which carries both hypercolor and color. Besides, there is also a massive $Z'_E$ boson which corresponds to the diagonal $U(1)_E$ subgroup of $SU(N+3)_{E}$. The generic charges of fermions under this broken $U(1)_E$ are given by
\begin{align}
{\psi_L},\,{\psi_R}: -{1}/{N},\quad  {q_L},\,{t_R} : {1}/{3}~,
\end{align}
which features a universal charge in the SM sector. This $Z'_E$ is the source of dangerous flavor processes such as flavor-changing neutral currents. However, if it is third-generation-philic, the flavor constraints are much weaker, which has been studied in \cite{Chung:2021ekz, Chung:2021xhd}.

In this type of extension, we can easily combine $ \mathcal{G}_{HC}$ and $\mathcal{G}_{SM}$ to $\mathcal{G}_E$ and thus hyperfermions and top quarks to multiplets $Q$. Since the $E_\mu$ boson carries hypercolor, it will form a hypercolor singlet bound state with other hypercolored particles below $\Lambda_{HC}\sim 10$ TeV. Hence, even if it has a mass as light as $1$ TeV, there won't be new states observable around the TeV scale, potentially explaining the absence of new particles so far. The only exception is $Z'_E$, which can be searched for at the LHC.

However, since the $SU(N)_{HC}$ group is directly separated from the $SU(N+3)_{E}$ group. The gauge coupling $g_{E}$ is the same as hypercolor coupling $g_{H}$ above the breaking scale $f_{E}$. After breaking, the running can separate the two couplings. However, to generate the observed top Yukawa $y_t\sim1$, the two scales $f_{E}$ and $f_{}$ must be close, which means $g_{E}$ must be close to $g_{H}$,  the strong hypercolor coupling. Therefore, the resulting $M_E\sim g_Ef_E$ is expected to be very heavy and the fine-tuning problem from the top loop will not be relieved.

\subsection{New extension: $\mathcal{G}_{HC} \times(\mathcal{G}_{HF} \times \mathcal{G}_{SM} \subset \mathcal{G}_{E})$}\label{sec:New}

The new extension will be the main focus of this study. To avoid a large $g_{E}$ situation as mentioned, we want to decouple it from the hypercolor coupling $g_H$. As all we need is to have hyperfermions and top quarks in the same representation, the unification of the gauge group is not necessary. One can imagine the combination happens in an orthogonal direction to the hypercolor group such that the couplings $g_{E}$ and $g_{H}$ are unrelated. In this case, the coupling $g_E$ is related to one of the SM gauge couplings instead. The gauge group $\mathcal{G}_{E}$ can be weakly coupled and is broken down to $\mathcal{G}_{HF} \times \mathcal{G}_{SM}$ at the scale $f_E$, where $HF$ stands for hyperfermion.

More specifically, we consider the extension of the SM $SU(3)_C$ to $SU(4)_{EC}$ to include the hyperfermions, where $EC$ stands for extended color. The fermion content under $SU(N)_{HC} \times SU(4)_{EC}\times  SU(2)_W$ is given by
\begin{align}
Q_L=(N,4,2),\quad Q_R=(N,4,1)~,
\end{align}
After the first breaking, the $SU(4)_{EC}$ gauge group is broken down to $SU(3)_{EC}$. The fermion content then becomes (under $SU(N)_{HC} \times SU(3)_{EC} \times SU(2)_W$)
\begin{align}
\text{Left-handed (LH): }&(N,3,2),~(N,1,2)~\nonumber\\
\text{Right-handed (RH): }&(N,3,1),~(N,1,1)~,
\end{align}
which should include both hyperfermions and top quarks. However, under this setup, all the fermion are charged under $SU(3)_{HC}$, which is obviously not allowed for a realistic top quark. Unless the $SU(N)_{HC}$ is broken and thus unconfined like Topcolor \cite{Hill:1991at}. The fact that top quarks are charged under hypercolor also restricts the number of $N$ we can have (unlike the traditional extension) because we can not introduce exotic degrees of freedom for top quarks. Instead, we can only use the existing quantum number in the SM top quark, such as $N=3$ in the Topcolor models, and have the SM gauge group as the unbroken subgroup through an additional breaking process $SU(N)_{HC} \times SU(N)_{ESM} \to SU(N)_{SM}$.

In general, we can have hypercolor group as $SU(3)_{HC}$ (broken down to $SU(3)_C$ in the end), $SU(2)_{HC}$ (broken down to $SU(2)_W$ in the end), or $U(1)_{HC}$ (broken down to $U(1)_Y$ in the end). In this work, we focus on the first case with $N=3$. Therefore, an additional breaking process is required to break $SU(3)_{HC} \times SU(3)_{EC} \to SU(3)_{C}$ and the fermion content is further separated to (under $SU(N)_{HC} \times SU(3)_{EC} \times SU(2)_W \to SU(3)_{C} \times SU(2)_W$)
\begin{align}
&Q_L \to (3,3,2)+(3,1,2) \to (6,2)+(\bar{3},2)+(3,2),\\
&Q_R \to (3,3,1)+(3,1,1) \to (6,1)+(\bar{3},1)+(3,1),
\end{align}
which includes exotic fermions transformed as sextets. For anti-triplets and triplets, though they look similar, they have different strengths of interactions as the anti-triplet originated from $(3,3)$ with both $SU(3)$ interactions but the triplet only has the one from $SU(3)_{HC}$. This difference is crucial to realize the tilting mechanism and requires the anti-triplets to be hyperfermions and triplets to be top quarks. Together with exotic fermions labelled by $f_{L,R}$, we get
\begin{align}
&Q_L\to~f_L=(6,2),\quad\psi_L=(\bar{3},2),\quad q_L=(3,2)~,\\
&Q_R\to~f_R=(6,1),\quad\psi_R=(\bar{3},1),\quad t_R=(3,1)~.
\end{align}
This setup can allow $\bar{\psi}\psi$ to form the condensate without $\bar{t}t$ condensate. Such a condition might require some fine-tuning among the couplings as in Topcolor models \cite{Hill:1991at}. But the self-breaking mechanism could fix the strong coupling at the value right above the critical point, which can make the tilting mechanism look natural. More concrete discussions will be presented in the next section.

In this type of extension, we can still combine hyperfermions and top quarks but through a more complicated way with a cost of exotic fermions. Also, the top quark now also undergoes the hypercolor interaction. However, the $E_\mu$ gauge boson no longer carries hypercolor and is naturally light, which can cut off the top loop below the TeV scale. There is still a massive $Z'_E$ boson which plays an important role in phenomenology.

%\newpage

\section{A concrete model}\label{sec:Model}

In this section, we construct a concrete model based on $SU(4)_{EC}$ with all the ingredients we mention. For the gauge sector, we consider a strongly coupled $SU(3)_{HC}$ and a weakly coupled $SU(4)_{EC}$. The overall gauge group is $\mathcal{G}_E=SU(3)_{HC}\times SU(4)_{EC}\times SU(2)_W \times U(1)_X$ \footnote{A similar group structure and breaking pattern has also been studied know as "4321 model" \cite{DiLuzio:2017vat} for the purpose of TeV-scale leptoquarks and B-meson anomalies.}. We denote the corresponding gauge fields as $H_\mu^{a}$, $E_\mu^\alpha$, $W_\mu^i$, and $X_\mu$, the gauge couplings as $g_H$, $g_E$, $g_W$, $g_X$, and the generators as $T^a$, $T^\alpha$, $T^i$, $Y'$ with indices $a=1,...,8$, $\alpha=1,...,15$, $i=1,2,3$. The generators are normalized as Tr$(T^AT^B)=\frac{1}{2}\delta^{AB}$.

The gauge group is spontaneously
broken down to SM gauge group $\mathcal{G}_{SM}=SU(3)_{C} \times SU(2)_W \times U(1)_Y$ through the scalar representation $\Sigma = (\bar{3},4,1,1/24)$, which acquires a vacuum expectation (VEV) value given by
\begin{equation}\label{ECVEV}
\langle \Sigma\rangle = \frac{f_{E}}{\sqrt{2}}
\begin{pmatrix}
0  &  1  &  0  &  0   \\
0  &  0  &  1  &  0   \\
0  &  0  &  0  &  1   \\
\end{pmatrix}.
\end{equation}
The formation of the $\Sigma$ field and its VEV can be realized dynamically through the tumbling gauge theory with additional chiral fermion under larger representation, which will be discussed in subsection \textbf{A}.

The breaking pattern of $\mathcal{G}_E\to \mathcal{G}_{SM}$ can be separated into three parts corresponding to the three resulting massive gauge bosons with different tasks:
\\
(1) $SU(4)_{EC}\to SU(3)_{EC}\times U(1)_{EC}$ breaking introduces the massive $E_\mu$ boson with the mass
\begin{align}
M_E=\frac{1}{2}\,g_Ef_{E}~
\end{align}
and the gauge coupling $g_E$. It plays an important role in connecting the hyperfermions with top quarks, which helps generate the top Yukawa coupling. The mass $M_E$ thus serves as the cutoff scale of top loop correction to the Higgs quadratic term.
\\
(2) $SU(3)_{HC}\times SU(3)_{EC} \to SU(3)_C$ breaking leads to a broken $SU(3)'$ and an unbroken $SU(3)$ expressed as
\begin{align}
{G'}_\mu^a=\frac{g_HH_\mu^a-g_EE_\mu^a}{\sqrt{g_H^2+g_E^2}},\quad
{G}_\mu^a=\frac{g_EH_\mu^a+g_HE_\mu^a}{\sqrt{g_H^2+g_E^2}}~,
\end{align}

The broken $SU(3)'$ bosons get the mass
\begin{align}
M_{G'}=\frac{1}{\sqrt{2}}\,\sqrt{g_H^2+g_E^2}f_{E}
\end{align}
and the gauge coupling $g'_s=\sqrt{g_H^2+g_E^2}$. 
It is the mediator of strong interaction and makes the hyperfermions condense, which leads to the subsequent composite Higgs and EWSB. More details are covered in subsection \textbf{B}.

The unbroken $SU(3)$ is just SM color group $SU(3)_C$ with the gauge coupling given by
\begin{align}
g_s=\frac{g_Hg_E}{\sqrt{g_H^2+g_E^2}}=1.02~,
\end{align}
where we choose the matching value at the scale of $2$ TeV. The matching then fixes the value $g_E\sim g_s = 1.02$ assuming $g_H\gg g_E$, which is related to the SM 
coupling and is weak as desired. The mass $M_E$ is then determined, which will be discussed further in subsection \textbf{C}.
\\
(3) $U(1)_{EC}\times U(1)_X \to U(1)_Y$ breaking similarly leads to a broken $U(1)'$ and an unbroken $U(1)$ expressed as
\begin{align}
Z'_{E,\mu}=\frac{cg_X X_\mu-g_EE_\mu^{15}}{\sqrt{c^2g_X^2+g_E^2}},~
{B}_\mu=\frac{g_EX_\mu+cg_XE_\mu^{15}}{\sqrt{c^2g_X^2+g_E^2}}~,
\end{align}
where $c=1/\sqrt{24}$. The $Z'_E$ boson gets the mass
\begin{align}
M_{Z'}=\frac{1}{\sqrt{8}}\,\sqrt{c^2g_X^2+g_E^2}\,f_{E}
\end{align}
and the gauge coupling gauge coupling $g'=\sqrt{c^2g_X^2+g_E^2}$. It is the lightest new degree of freedom and has a huge impact on phenomenology. 

The unbroken $U(1)$ would be the SM hypercharge with $Y=c\,T^{15}+X$ where $T^{15}=1/\sqrt{24}\,\text{diag}(3,-1,-1,-1)$. The gauge coupling is given by
\begin{align}
g_Y=\frac{g_Xg_E}{\sqrt{c^2g_X^2+g_E^2}}=0.36~,
\end{align}
where we choose the matching value at the scale of $2$ TeV. The matching then fixes the value $g_X\sim g_Y =0.36$ because $g_E\sim 1.02$ is much greater than $cg_Y$.

Based on the matching with SM gauge coupling, we get the strengths of new gauge groups within $\mathcal{G}_E$ as
\begin{align}
g_E\sim g_s = 1.02,\quad g_X\sim g_Y =0.36~.
\end{align}
The strong coupling $g_H$ is expected to be right below the critical coupling $g_c \sim 5.1$ which will be explained in subsection \textbf{B}.

Next, we discuss the fermion content. In this part, we only focus on the relevant content for the generation of the top (and bottom) Yukawa coupling. Additional fermions might be added to realize the tumbling mechanism or to get a realistic composite Higgs sector, which will be discussed in subsection \textbf{A} and \textbf{B}. Besides, We remain agnostic about how the other light SM fermions obtain their masses and assume that the required mechanisms are separated from our current work and do not worsen the hierarchy problem, which could be true due to their small Yukawa couplings.\footnote{The separation can be realized in a family non-universal extension of the SM gauge group, such as in \cite{Hill:1991at,DiLuzio:2017vat}. The detailed construction is beyond the scope of this study and we leave it to the future study.} Therefore, we will only address the SM third generation quark - the top and bottom quark, especially on the top quark, in the following discussion.

The required fermions under $SU(3)_{HC}\times SU(4)_{EC} \times SU(2)_W\times U(1)_X$ are given by 
\begin{align}
Q_L=(3,4,2,\frac{1}{24}),\quad U_R/D_R=(3,4,1,\frac{1}{24}\pm \frac{1}{2})~.
\end{align}
The extension is anomaly-free under the gauge groups except that there is Witten anomaly. The problem can be solved with one additional $SU(2)$ doublet fermion, which is chargeless under $U(1)_X$. Since it doesn't carry $U(1)$ charge, we can decouple it by writing down a Majorana mass term without breaking any gauge symmetry.

After the symmetry breaking, the fermions are decomposed as (under $\mathcal{G}_{SM}=SU(3)_{C}\times SU(2)_W\times U(1)_Y$)
\begin{align}
Q_L&\to (6,2)_0+(\bar{3},2)_0+(3,2)_\frac{1}{6}~,\\
U_R&\to (6,1)_\frac{1}{2}+(\bar{3},1)_\frac{1}{2}+(3,1)_\frac{2}{3}~,\\
D_R&\to (6,1)_{-\frac{1}{2}}+(\bar{3},1)_{-\frac{1}{2}}+(3,1)_{-\frac{1}{3}}~.
\end{align}

Each of fermion multiplets is separated to three parts, exotic fermions $f_{L,R}$, hyperfermions $\psi_{L,R}$, and the SM quarks $q_L, t_R, b_R$ as
\begin{align}\label{fermions1}
&f_L=(6,2)_0,~  \psi_L = (\bar{3},2)_0,~q_{L} = (3,2)_\frac{1}{6}~,\\
&f_{U,R}=(6,1)_\frac{1}{2},~\psi_{U,R}=(\bar{3},1)_\frac{1}{2},~t_{R} = (3,1)_\frac{2}{3}~,\label{fermions2}\\
&f_{D,R}=(6,1)_{-\frac{1}{2}},~\psi_{D,R}=(\bar{3},1)_{-\frac{1}{2}},~b_{R} = (3,1)_{-\frac{1}{3}}~.\label{fermions3}
\end{align}

In the following three subsections, we will discuss the roles of each fermion and all the relevant mechanisms from the top down in order of energy scales as \\
\textbf{A.}  The $\mathcal{G}_E \to \mathcal{G}_{SM}$ breaking at the scale $f_{E}\sim 1.7$ TeV through tumbling mechanism with exotic fermions  \\
\textbf{B.}  Composite Higgs formation at the scale $f\sim 1$ TeV through hyperfermion condensation \\
\textbf{C.}  Generation of top Yukawa coupling at the scale $M_{E}$ through integrating out the $E_\mu$ boson\\
Then we summarize the overall spectrum and properties of new particles in subsection \textbf{D}.

%\newpage

\subsection{Tumbling mechanism with exotic fermions }\label{sec:Tumbling}

In this model, the first symmetry breaking required is $SU(3)_{HC}\times SU(4)_{EC} \to SU(3)_C$, which is similar to the 4321 model \cite{DiLuzio:2017vat}. Besides using additional scalars with nonzero VEVs to realize the breaking, we would like to construct a dynamical model with the breaking through the $SU(3)_{HC}$ strong interaction itself. Such self-breaking mechanism is known as "Tumbling" gauge theories \cite{Raby:1979my} and has been used in BSM model building \cite{Martin:1992aq, Martin:1992mj}.

The self-breaking of strong $SU(3)$ gauge group has already been studied in \cite{Amati:1980wx,Gusynin:1982kp} and the desired breaking is possible in a chiral theory with fermions in both the triplet $\bf{3}$ and sextet $\bf{6}$ representation. Since we already have LH $\bf{3}$, we only need to add an additional RH $\bf{6}$. With fermions under $\mathcal{G}_E$ given by
\begin{align}
Q_L=(3,4,2,{1}/{24})~,\quad F_R=({6},1,{2},0),
\end{align}
the most attractive channel under $SU(3)_{HC}$ is RH $\bf{6}$ combined with some of LH $\bf{3}$ to form the condensate. The $SU(3)_{HC}$ will be broken down to a $SU(3)$ symmetry which is the diagonal subgroup of $SU(3)_{HC}\times SU(3)_G$, where $SU(3)_G$ is a subgroup of global symmetry of $\bf{3}$. The global symmetry of $\bf{3}$ under our setup will be the $SU(4)_{EC}\times U(1)_X$ gauge symmetry. The $SU(2)_W$ part is directly contracted so it does not play any role here. The condensate, $\bar{F}_RQ_L$, is formed with exactly the same quantum number as the scalar $\Sigma = (\bar{3},4,1,1/24)$ and with the desired VEV structure shown in Eq. \eqref{ECVEV}. The scale $f_E$ is determined by the strength of $\bf{\bar{6}\,3}$ condensate and the coupling $g_H$ is fixed at the corresponding value.

The VEV not only breaks $SU(3)_{HC}\times SU(4)_{EC}\times U(1)_X$ to $SU(3)_C\times U(1)_Y$ with massive $E_\mu$, $G'$ and $Z'_E$ but also gives the Dirac masses to the fermion sextet. The VEV mixes the $F_R$ with the exotic fermion $f_L$ in Eq. \eqref{fermions1}. We then get the mass term as $M_F\bar{F}_Rf_L$ with $M_F\sim Y_{S}f_{E}$, where the Yukawa coupling $Y_{S}$ comes from the strong dynamics and should have a large value. With the assistance of the tumbling mechanism, now we have a dynamical origin for the breaking pattern and also get rid of part of the dangerous exotic fermions as they are much heavier and out of reach of LHC searches.

Similarly, we can introduce two additional LH sextets $F_{U,L}$ and $F_{D,L}$ to generate Dirac masses with $f_{U,R}$ and $f_{D,R}$. However, an additional mechanism is required to forbid the direct condensation among the sextets $F_L$ and $F_R$, which is more attractive, such as a strong repulsive $U(1)$ force. Moreover, two additional fermion sextets will flip the sign of the hypercolor's beta function, which will ruin the whole strong dynamics. Therefore, to realize the tumbling mechanism with an anomaly-free fermion content, a more complicated fermion content is required but we leave it for the future study.

%\newpage

\subsection{Composite Higgs from hyperfermion condensate}\label{sec:CHM}

After the first breaking, the strong $SU(3)_{HC}$ is broken and the fermion sextets become massive. The next most attractive channels are RH $\bf{3}$ combined with LH $\bf{3}$ whose strength of the attraction is only slightly below the first one \cite{Martin:1992aq, Martin:1992mj}. Though $SU(3)_{HC}$ is already broken and the coupling $g_H$ is fixed at the value to trigger $\bar{F}_RQ_L=\bf{\bar{6}\,3}$ condensation, we assume the $\bar{\psi}_R\psi_L=\bf{\bar{3}\,3}$ condensate can still happens with an assist from $SU(3)_{EC}$ interaction.

Since the strong gauge group is broken, we can describe it by the Nambu-Jona-Lasinio (NJL) model \cite{Nambu:1961tp, Nambu:1961fr}. The critical coupling for $\bf{\bar{3}\,3}$ condensation is
\begin{align}
g_c=\sqrt{{8\pi^2}/{3}}\sim 5.1~.
\end{align}
We claim that after the first breaking, the coupling $g_H$ is fixed at the value right below $g_c$ as the first attractive channel with $\bf{\bar{6}\,3}$ has a smaller critical coupling.

Combining with the $SU(3)_{EC}$ interaction, which only applies on hyperfermions but not top quarks, we claim the following relation on couplings is achieved
\begin{align}
g_\psi^2 \sim g_H^2+g_E^2>g_c^2\,,\quad g_t^2 \sim g_H^2<g_c^2~,
\end{align}
such that the interaction is strong enough to form $\bar{\psi}\psi$ condensate for composite Higgs without $\bar{t}t$ condensate.

In the NJL model, we can also estimate the breaking scale by the $\bar{\psi}\psi$ condensate. Generically, the breaking scale in the NJL model is close to the scale of the broken strong gauge group, i.e. $f \sim f_E$, unless we have $g_\psi \sim g_c$. However, as we already show how the coupling $g_\psi$ can be naturally closed to critical coupling $g_c$ in our model, we can then get a desired hierarchy $f < f_E$. The difference thus determines the value of $y_t$ in the model.

The detail of the composite Higgs sector is model-dependent as the Higgs could be pion-like resonance in composite Higgs models (CHM) or sigma-like resonance in technicolor models (TC). Use the former one as an example. In the fundamental composite Higgs models (FCHM), we need the $\bar{\psi}\psi$ condensate to break the global symmetry at the breaking scale $f$ and introduce Higgs as pseudo-Nambu-Goldstone bosons (pNGBs) of the coset. With $SU(3)_{HC}$ strong interaction and the hyperfermions under complex representations, the minimal choice of the FCHMs \cite{Cacciapaglia:2014uja,Cacciapaglia:2020kgq} is the one with four Dirac fermions in the (anti-)fundamental representation, which results in a $SU(4)\times SU(4)/ SU(4)$ FCHM. The quantum numbers of Dirac hyperfermions are given by
\begin{align}\label{4fermions}
\Psi_{1} = (\bar{3},2)_0,~\Psi_{2} = (\bar{3},1)_\frac{1}{2},~\Psi_{3} = (\bar{3},1)_{-\frac{1}{2}},
\end{align}
where we use $\bar{3}$ instead of ${3}$ to match our fermion content. Compared to Eq. \eqref{fermions1}-\eqref{fermions3}, we find the $\psi_{L}$, $\psi_{U,R}$, and $\psi_{D,R}$ match the required fermions $\Psi_{1,L}$, $\Psi_{2,R}$, and $\Psi_{3,R}$, which are the fermion components of the composite Higgs. The complete model should contain eight Weyl hyperfermions so the fermion content should be extended with four more hyperfermions of desired quantum numbers to ensure the formation of electroweak preserving condensate, which is the main difference between the composite Higgs models and the technicolor models. On the other hand, additional fermions might not be required if one can realize the idea in the technicolor models. In the following discussion, we will have a pNGB Higgs in our mind.

In general, there should be two Higgs doublets with $H_1 \sim \bar{\psi}_{U,R}{\psi}_{L}$ and $H_2 \sim \bar{\psi}_{D,R}{\psi}_{L}$. We expect the $H_1$ being the SM-like Higgs and $H_2$ being a heavy second Higgs doublet. Since the goal of this study is to generate top Yukawa coupling, we will not dig into the details of the composite Higgs sector but refer the readers to other dedicated studies of this type of FCHMs \cite{Ma:2015gra, Wu:2017iji}.

%\newpage
\subsection{Top Yukawa from the $E_\mu$ boson}\label{sec:ytfromEC}

The top Yukawa model that we construct through the extended gauge group $SU(4)_{EC}$ introduces top Yukawa coupling in exactly the way we describe in Sec. \ref{sec:Basic}. Now with a concrete model, we can further estimate the required value and set up our benchmark.

With the extended gauge group $SU(4)_{EC}$ broken at the scale $f_{E}$, the $E_\mu$ gauge boson which connects the hyperfermions and top quarks acquires a mass $M_E$. The top Yukawa coupling is generated after the composite Higgs is formed by the $\bar{\psi}_{U,R}{\psi}_{L}$ condensate and the massive $E_\mu$ boson is integrated out. The value is given by
\begin{align}
y_t\sim \frac{1}{v}\frac{g_{E}^2}{M_{E}^2}\langle \bar{{\psi}}_{U,R}{\psi}_{L}\rangle_{HC}
\sim  \left(\frac{f_{}}{f_{E}}\right)^2Y_{S}~,
\end{align}
where $Y_{S}$ is the Yukawa coupling from the strong interaction among hyperfermions. In the NJL model, $Y_s$ can be estimated as 
\begin{align}
Y_{S}\sim \frac{4\pi}{\sqrt{N_{HC}\text{ ln}(\Lambda^2/M_\psi^2)}}~,
\end{align}
where $\Lambda$ is the cutoff of the theory and $M_\psi$ is the dynamical mass of hyperfermions. In a strongly coupled theory, $Y_{S}$ is expected to be $3-4$. In our case, as we have additional splitting between $f_E$ and $f$ which might enhance the logarithmic term, we take the lower value $Y_s=3$ for our numerical study.

To generate the observed top Yukawa $y_t \sim 1$, the scale 
\begin{align}
f_{E}\sim \sqrt{Y_{S}} \times f \sim 1.7 \times f~.
\end{align}
Setting $f=1$ TeV as our benchmark, we get $f_{E}\sim 1.7$ TeV. Next, we can also derive the mass of the $E_\mu$ gauge boson. With $g_E\sim 1.02$ and $f_{E}\sim 1.7$ TeV, The mass is then given by
\begin{align}
M_E=\frac{1}{2}\,g_Ef_{E}\sim 0.9~\text{TeV}~,
\end{align}
which is the most important quantity in our model because it serves as the cutoff of the top loop. That is, the top Yukawa coupling is only generated below the scale of $M_E\sim 0.9~\text{TeV}$, where the $E_\mu$ gauge boson is integrated out. When approaching the mass $M_E$, the top Yukawa coupling will start revealing its original nature as
\begin{equation}\label{ytUV}
y_t(k^2) \sim \frac{y_{t,0}}{(1+k^2/M_E^2)}\,,
\end{equation}
where $k$ is the momentum related to the vertex and  $y_{t,0}$ is the top Yukawa coupling at $k^2=0$. One can substitute the modified top Yukawa coupling above into Eq.~\eqref{toploop0}. The resulting top loop contribution becomes
\begin{align}\label{toploopUV}
\Delta m_H^2|_{\text{top}}
\sim-i\,2N_c \int_{}^{} \frac{d^4k}{(2\pi)^4}y_t^2(k^2) \frac{1}{k^2}
= -\frac{3}{8\pi^2}y_{t,0}^2 M_{E}^2~,
\end{align}
where the mass $M_E$ now plays the role of $\Lambda_t$ as it supposed to be. With the weakly coupling extended gauge group $SU(4)_{EC}$, we then get a naturally light cutoff for the top loop contribution, which can relieve the fine-tuning problem and serve as a good alternative to the top partner solution.

Notice that, a similar Yukawa coupling term for the bottom quarks with $y_b\sim 1$ will also be generated but with the second Higgs $H_2$. Such a term, if contributing to all the bottom quark mass, will lead to a generic Type-II two Higgs doublet model with a large tan$\beta$. However, the bottom quark mass can also come from the top quark mass through other mechanisms such as radiative mass generation \cite{He:1989er, Ma:1990ce, Babu:1990vx, Dobrescu:2008sz, Baker:2020vkh}. Since the low-scale bottom Yukawa coupling is not a necessary part of the model, one can even replace the $D_R$ in the fermion content such that the bottom Yukawa will not be generated at the tree level, such as in \cite{Cacciapaglia:2015yra}. Due to this freedom, we will only focus on the new particles that are relevant for the generation of top Yukawa coupling in the following discussion.

%\newpage

\subsection{The overall spectrum}\label{sec:Spectrum}

Before moving on to the phenomenology section, we briefly summarize all the relevant new particles we introduce and the overall spectrum. Start with massive gauge bosons, we have the broken $SU(3)'$ bosons $G'_\mu$, which is a color-octet (colorons) with masses given by
\begin{align}
M_{G'}=\frac{1}{\sqrt{2}}\,\sqrt{g_H^2+g_E^2}f_{E}\sim 6 \text{ TeV}~.
\end{align}
Next, the $E_\mu$ gauge boson, with quantum number under $\mathcal{G}_{SM}$ as $(3,1,-1/6)$, is much lighter with a mass
\begin{align}
M_E=\frac{1}{2}\,g_Ef_{E}\sim 0.9~\text{TeV}~.
\end{align}
Last, there is a massive neutral bosons $Z'_E$ with a mass
\begin{align}
M_{Z'}=\frac{1}{\sqrt{8}}\,\sqrt{c^2g_X^2+g_E^2}\,f_{E}\sim 0.6~\text{TeV}~,
\end{align}
which is the lightest new particles. As $g_E\gg cg_X$, the couplings between $Z'_E$ and fermions are mainly determined by the $U(1)_{EC}$ part with coupling $g_E$ and charge of fermions given by
\begin{align}\label{Z'_EQ}
q_L,\,t_R:\frac{3}{\sqrt{24}}\sim 0.6,\quad \psi_{L,R},\,f_{L,R}:\frac{-1}{\sqrt{24}}\sim -0.2~,
\end{align}

Besides bosons, we have some new fermions at the TeV scale. Because the $SU(3)_{HC}$ is broken, these fermions are unconfined and can be searched for at the LHC. First, we have color sextet Dirac fermions $F$ with quantum number $(6,2,0)$ and $(6,1,\pm 1/2)$, which get a dynamical mass at the breaking scale $f_E$ with \footnote{Here we still use $Y_s=3$ for convenience. However, for a sextet fermion, due to a stronger interaction, the coupling $Y_s$ should be greater and the sextet fermions $F$ should be heavier.}
\begin{align}
M_{F}\sim Y_s f_{E}\sim 5~\text{TeV}~ .
\end{align}
Next, the hyperfermions $\psi$ are also unconfined. They are also Dirac fermion with quantum number $(\bar{3},2,0)$ and $(\bar{3},1,\pm 1/2)$. The mass is lighter as it comes from a lower breaking scale $f$ as
\begin{align}
M_{\psi}\sim Y_s f_{}\sim 3~\text{TeV}~.
\end{align}

%\newpage

\section{INDIRECT SEARCHES}\label{sec:Indirect}

Since the goal of the whole study is to generate the top Yukawa coupling, we will start with the discussion on top physics. The main effect comes from the dimension-six nature of top Yukawa coupling, which has already been discussed in \cite{Bally:2022naz,Bally:2023lji}, so in this paper, we will focus on the benchmark we use and some new analyses.

\subsection{Higgs coupling measurements} \label{sec:kappa}

Having the top Yukawa from dimension-six operators in general will not affect its value $y_{t,0}$ at zero momentum. However, a deviation is still expected due to the Goldstone nature of Higgs in CHMs. The measurements of the top Yukawa coupling as well as other Higgs couplings are the direct test of misalignment, which is the key mechanism in CHMs. Combining all the Higgs coupling measurements, we can get a constraint on the breaking scale $f$. Assuming a simplified form with $\kappa_V=\kappa_f=\sqrt{1-v^2/f^2}$ for the deviations on the Higgs couplings, recent measurements by ATLAS and CMS with Run 2 data \cite{ATLAS:2022vkf, CMS:2022dwd} put a constraint on the scale $f > 1.1$ TeV, which is slightly above our benchmark with $f = 1$ TeV. The constraint can be relieved if we go beyond the simplified form.

\subsection{Running Top Yukawa}\label{sec:ytrun}

The dimension-six origin of the top Yukawa coupling will lead to a nontrivial form factor on the top-Higgs vertex. Such momentum-dependence of the top Yukawa coupling at high scales could be measured in the tails of momentum distributions in processes such as $t\bar{t}h$ production \cite{Goncalves:2018pkt, Goncalves:2020vyn, MammenAbraham:2021ssc, Bittar:2022wgb}. 
However, it will require precise measurement of $t\bar{t}h$ differential cross section, which suffers from both the small $t\bar{t}h$ cross section and the complexity of final states. The current measurement has not yet reached the desired sensitivity but could be done with new data at the HL-LHC.

\subsection{Running Top mass}\label{sec:mtrun}

\begin{figure}[tbp]
\centering
\includegraphics[width=0.48\textwidth]{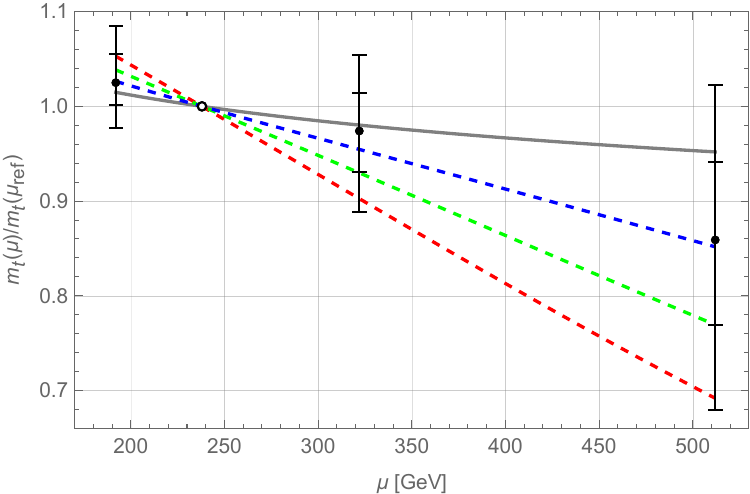}
\caption{The top mass running in the SM (gray) v.s. the running in Eq. \eqref{ytmu2} with $\Lambda_t=700$ GeV(red), $900$ GeV(green), and $1300$ GeV(blue) compared with the data points from \cite{Defranchis:2022nqb} (the inner bars represent $1\sigma$ uncertainties and the outer bars for $2\sigma$ uncertainties). \label{mtrunning}}
\end{figure}

Tests of the dimension-six top Yukawa can also be done by measuring the running of the top quark mass. The nontrivial running top mass at the high scale will affect the $t\bar{t}$ differential cross section. Compared to the $t\bar{t}h$ channel, the $t\bar{t}$ channel has a larger cross section, which could provide better sensitivity. The measurement has been done by the CMS collaboration using part of Run 2 data with an integrated luminosity of 35.9 fb$^{-1}$ \cite{CMS:2019jul}. The result has been interpreted in \cite{Defranchis:2022nqb} as the top mass running up to $0.5$ TeV as shown in Fig. \ref{mtrunning}.

Assume a generic form of top mass running as
\begin{align}
m_t(\mu)=m_{t,SM}(\mu)\left(\frac{\Lambda_t^2}{\mu^2+\Lambda^2_t}\right)~,
\label{ytmu2}
\end{align}
where $\Lambda_t=M_{E}$ in our top Yukawa model. We can then get a bound from the current data as $M_{E}\gtrsim 700$ GeV. With more data coming out, we expect the relevant parameter space can be fully explored in the HL-LHC era.

\subsection{Four top quarks cross section}\label{sec:4tpheno}

The model also comes with new bosons interacting with top quarks, including a massive neutral boson $Z'_E$ and colorons $G'$. Both of them will introduce additional contributions to the four top-quark cross section. Due to the heaviness of top quarks, this measurement is like a precision test of a rare process. In the SM, the cross section is derived as \cite{vanBeekveld:2022hty}
\begin{align}
\sigma_{t\bar{t}t\bar{t}}^{\text{SM}}=13.4^{+1.0}_{-1.8} \text{ fb.}
\end{align}

Measurements using different final states have been performed by both ATLAS \cite{ATLAS:2023ajo} and CMS \cite{CMS:2023ica} with LHC Run 2 data. The cross section are measured as
\begin{align}
\sigma_{t\bar{t}t\bar{t}}^{\text{ATLAS}}=22.5^{+6.6}_{-5.5} \text{ fb,}\quad
\sigma_{t\bar{t}t\bar{t}}^{\text{CMS}}=17.9^{+4.4}_{-4.1} \text{ fb.}
\end{align}
where ATLAS gets a central value of about $1.7$ times the SM prediction while CMS gets a value closer to the SM prediction but still higher.

Both collaborations have seen evidence for the simultaneous production of four top quarks and a cross section slightly larger compared to the SM prediction. The bound on the cross section at 95\% CL level is given by
\begin{align}
\sigma_{t\bar{t}t\bar{t}}^{\text{}}<38\,(27) \text{ fb   from ATLAS (CMS).}
\end{align}
Several analyses aiming at four top final states have been performed in recent years \cite{Darme:2021gtt, Banelli:2020iau, Blekman:2022jag}. Following the analysis of simplified models in \cite{Darme:2021gtt}, we get a constraints on the ratio between the mass $M_V$ and coupling $g_V$ of a top-philic vector color-singlet boson ($Z'$) as
\begin{align}
\frac{M_V}{g_V}> 0.48\,(0.56) \text{ TeV   from ATLAS (CMS),}
\end{align}
or the mass $M_C$ and coupling $g_C$ of a top-philic vector color-octet boson (coloron) as
\begin{align}
\frac{M_C}{g_C}> 0.35\,(0.40) \text{ TeV   from ATLAS (CMS).}
\end{align}
The coloron's contribution is only weakly constrained and should be subleading. The main contribution is from the $Z'_E$ where we have $g_V\sim g_E\times 3/\sqrt{24} \sim 0.6$. It is below the current constraint and could provide an explanation for the observed excess.

\subsection{Flavor constraints}\label{sec:Flavor}

Besides the four top quarks cross section, the same four-quark operators will also introduce other light quark physics through the mixing, which might lead to dangerous flavor-changing neutral currents. Assuming that the mixing angle $\theta_{23}\gg\theta_{13}$ analogous to the CKM matrix, then among all the processes, the strongest constraint comes from $B_s-\bar{B}_s$ mixing, which contains both the second and third generation down-type quarks. The contribution comes from the operator
\begin{equation}
\Delta\mathcal{L}_{B_s}=C_{sb}(\bar{s}_L\gamma_\mu b_L)(\bar{s}_L\gamma_\mu b_L).
\end{equation}
Following the calculation in \cite{DiLuzio:2017fdq}, we can derive the contribution from new physics on the mass difference $\Delta M_s$. Comparing the measurement of mixing parameter \cite{HFLAV:2019otj} to the SM prediction by sum rule calculations \cite{King:2019lal}, we get the bound on the coefficient of the operator as
\begin{equation}\label{Bsmixing}
|C_{sb}| \approx \frac{1}{2}\left(\frac{g_{V}\theta_{sb}}{M_{V}\text{(TeV)}}\right)^2 \leq \left(\frac{1}{274}\right)^2
\end{equation}
for a top-philic color-singlet vector boson $V$, where the angle $\theta_{sb}$ is the mixing between the left-handed strange quark and bottom quark. In our benchmark with the ratio $g_V/M_V\text{(TeV)}=1$, we get a constraint on the mixing angle as $\theta_{sb} < 0.005$ which requires a special flavor structure for the down quark sector.

\subsection{Electroweak precision tests}\label{sec:EWPT}

Precise measurements from the electroweak sector typically impose strong constraints on new physics at the TeV scale, particularly concerning the $T$ parameter and $Zb\bar{b}$ coupling. Both of them measure the violation of $SU(2)_R$ symmetry, which is related to the detail of the second Higgs and the bottom Yukawa coupling. Since we only focus on the origin of the top Yukawa coupling, the topic is beyond the scope of this study and relies on the complete model with detailed study on the composite Higgs sector, which should preserve custodial symmetry to avoid strong constraints. We leave such a custodial symmetric model and the discussion of corresponding constraints for the future study.

%\newpage

\section{Direct searches}\label{sec:Direct}

There are many new particles in this top Yukawa model for the composite Higgs. Some of them are composite states from the strong sector but we will not discuss them. Instead, we would like to focus on the new particle due to the extension of the gauge group, including the new gauge bosons and fermions discussed in Sec. \ref{sec:Spectrum}.

\subsubsection{The $Z'_E$ boson}

Start from the lightest particle in the spectrum. First, there is a massive neutral boson $Z'_E$ with the mass $M_{Z'}\sim 600$ GeV. If the charge assignment follows only the Eq. \eqref{Z'_EQ}, the $Z'_E$ boson will only couple to the top and bottom quarks among the SM fermions. The dominant production will be through the $b\bar{b}$ fusion. In our model, there are only two decay channels with the final states $t\bar{t}$ and $b\bar{b}$. However, the current direct searches for both $t\bar{t}$ \cite{ATLAS:2023taw} and $b\bar{b}$ \cite{ATLAS:2019fgd} final states have no access to $M_{Z'}$ around $600$ GeV due to the heaviness of top quarks and the $b$-tagging issues.

Therefore, the direct search of $Z'_E$ is only possible with lepton final states but it requires an additional setup. Assuming that in a more realistic model, the $\tau$ lepton is also charged under $U(1)'$, then the most promising channel will become the process $b\bar{b}\to Z'_E \to \tau\tau$, which covers the sub-TeV regime. The current searches \cite{ATLAS:2020zms, CMS:2022goy} for $M_{Z'}=600$ GeV already require the cross section to be lower than $20$ fb which can put the constraint on the coupling of $Z'_E$ with $\tau$ leptons.

\subsubsection{The $E_\mu$ gauge boson}

Next, the $E_\mu$ boson is also at the sub-TeV scale with $M_{E}\sim 900$ GeV in our benchmark. The most important feature of $E_\mu$ is that it is stable! Since the hyperfermions get masses at few-TeV, without additional assumptions, there is no allowed decay channels for a single $E_\mu$ boson. It is the direct consequence of having a light mediator in an ETC-type model.

Under our extended gauge group, the $E_\mu$ is colored, which then gets a large cross section from the pair production process at the LHC. Although a single $E_\mu$ boson is stable, a pair of $E_\mu$ bosons is another story. For $pp \to E_\mu^+E_\mu^-$ at the LHC, both of them can decay to a top/bottom quark and an off-shell hyperfermion. In general, the off-shell hyperfermions can not decay to lighter on-shell final states so the $E_\mu$ boson is stable. However, the two off-shell hyperfermions from $E_\mu^+$ and $E_\mu^-$ experience a strong attraction which allow them to form a deeply bound state, which is just the composite Higgs in our model. Therefore, direct searches of an $E_\mu$ boson are unlikely in the LHC, but instead a BSM operator $\mathcal{O}_{tG}$ is generated, where
\begin{equation}\label{OtG}
\mathcal{O}_{tG}=g_s\left(\bar{q}_L\sigma^{\mu\nu}T_At_R\right)\tilde{H}G^A_{\mu\nu}+\text{h.c. .}
\end{equation}
The coefficient $C_{tG}$ of the operator, after integrating out the loop with $E_\mu$ bosons and hyperfermions, is given by
\begin{equation}\label{CtG}
C_{tG}\sim \frac{3}{16\pi^2}\frac{g_E^2Y_s}{M_\psi^2}\sim 0.007 \text{ TeV}^{-2}~
\end{equation}
using our benchmark value. The experimental constraint on the coefficient is analyzed using the $\bar{t}t$ final states measured by the CMS with part of run 2 data \cite{CMS:2019nrx} as
\begin{equation}\label{CtGexp}
-0.24 \text{ TeV}^{-2} <C_{tG}< 0.07 \text{ TeV}^{-2}
\end{equation}
at 95\% confidence level. Yet the constraint is an order of magnitude greater than the benchmark value because in our model the coefficient $C_{tG}$ is generated at the one-loop level but the desired precision can be reached in the near future. Besides, we also expect other interesting processes such as $E_\mu^+E_\mu^-\to t\bar{t}h\,/\,t\bar{t}Z\,/\,tbW$, which will affect the corresponding cross sections and can be tested in the HL-LHC era.

Strong constraints could arise from cosmology because it might introduce unacceptable relic abundance. Since the $E_\mu$ boson is stable and colored, it will form heavy color-neutral bound states with other colored particles through QCD interactions, which behave like massive stable charged particles. The constraints on stable charged particles have been studied \cite{Byrne:2002ri, Langacker:2011db}, which mainly depends on the thermal production/annihilation rate. Since the $E_\mu$ boson is colored, the relic abundance is lower compared to pure charged particles \cite{Wolfram:1978gp}. However, it also relies on the details of cosmological evolution such as reheating, so after all we only refer to the searches from the LHC.

\subsubsection{The $G'$ boson (Coloron)}

Compared to the $Z'_E$ boson and $E_\mu$ boson, the coloron is much heavier with the mass $\sim 6$ TeV. As a color-octet, we expect a large cross section even though it is heavy. If it only couples to the top and bottom quarks like the $Z'_E$ boson, the decay channels will also be dominated by $t\bar{t}$ and $b\bar{b}$ final states. However, due to the strong coupling $g'_s\sim g_H \sim 5$, we expect the coloron to be a very broad resonance, which will be hard to search for.

\subsubsection{Heavy fermions}

Since the strong $SU(3)_{HC}$ is broken and unconfined, new fermions, even charged under hypercolor, are able to propagate freely after being produced. There are two types of heavy fermions. First is the color sextet Dirac fermion $F$ with quantum number $(6,2,0)$ and $(6,1,1/2)$, which get a dynamical mass at the breaking scale $f_E$ with $M_{F}\sim 5$ TeV. Next, the hyperfermions $\psi$ are also Dirac fermion with quantum number $(\bar{3},2,0)$ and $(\bar{3},1,1/2)$, which have a lighter mass $M_{\psi}=3$ TeV. Both of them can be pair-produced at the LHC and decay through a $E_\mu$ boson plus a top/bottom quark channel. And again a pair of $E_\mu$ bosons will decay to two more top/bottom quarks with a Higgs/W/Z boson.

%\newpage

\section{Conclusion and Outlook} \label{sec:Conclusion}

In this paper, we study a Top Yukawa model based on the motivation from the Naturalness principle, i.e. a light cutoff for the top quark loop. We construct a composite Higgs model where the top Yukawa coupling arises from four-fermion interactions through an ETC-like mechanism. Different from the traditional extension, we extend the gauge group in a direction independent of the strong interaction. In this way, the gauge coupling $g_E$ can be weak and the mediator $E_\mu$, which plays the role of top loop cutoff, can be naturally light, which relieves the top loop contribution.

A concrete model with $\mathcal{G}_E=SU(3)_{HC}\times SU(4)_{EC}\times SU(2)_W \times U(1)_X$ is discussed in detail. The breaking of $\mathcal{G}_E\to \mathcal{G}_{SM}$ is realized dynamically through the tumbling mechanism with exotic chiral fermions. We also show that under this content, the hyperfermions can condense without the dangerous top quark condensation due to the tilting mechanism. Most important of all, the top Yukawa coupling is generated through a light mediator - the $E_\mu$ gauge boson from the weakly coupled $SU(4)_{EC}$ extended color group.

The rich phenomenology on top physics is discussed, where $t\bar{t}$ differential cross section could provide important hints. The method also features two new sub-TeV particles which have important impacts at the LHC. One is a third-generation-philic $Z'_E$ boson, the lightest state in the spectrum, which will enhance the $t\bar{t}t\bar{t}$ cross section. The other is the $E_\mu$ gauge boson, the cutoff of top loop, which will affect several final states with top quarks through a BSM operator $\mathcal{O}_{tG}$.

% outlook

This study aims at the model building in a different direction compared to the traditional model. Our attempt only focuses on the gauge group extension in its simplest way, which might not be realistic considering that we ignore the bottom quarks and other light fermions. We expect this extension can be applied to other flavor-safe setups, such as partial compositeness~\cite{Kaplan:1991dc}. In fundamental partial compositeness~\cite{Barnard:2013zea, Ferretti:2013kya}, the mixing should also arise from similar dimension-six four-fermion operators. With assistance from our method, the top partners no longer need to be light and can escape from the LHC direct searches without worsening the fine-tuning problem because now the top loop contribution is controlled by the light $E_\mu$ gauge boson~\cite{Chung:2024}.

Also, the detail of the composite Higgs sector is left aside to avoid distracting the attention from our goal. Because of the $SU(3)_{HC}$ hypercolor group, a large coset is expected. If we want to stick to the small coset, such as the $SU(4)/Sp(4)$ fundamental composite Higgs model with only a Higgs doublet and a real singlet \cite{Cacciapaglia:2014uja,Cacciapaglia:2020kgq}, we need hyperfermions to be pseudo-real representations of the hypercolor group. To realize the idea, we can have $SU(2)_{HC}$ with hyperfermions as fundamental representations, which is possible if the $SU(2)_{HC}$ is broken down to the SM $SU(2)_W$ in the end. For this scenario, we start with a strongly coupled $SU(2)_{HC}$ and a weakly coupled $SU(3)_{ESM}$. With a certain fermion content, we could have symmetry breaking $SU(3)_{ESM} \times SU(2)_{HC} \to SU(2)_W$ as desired. The concrete construction is left for future study.

Together with \cite{Bally:2022naz, Bally:2023lji}, we hope to raise some interest in the modified top Yukawa running scenario compared to the top partner solutions. As the constraints on the top partner mass become higher and require more fine-tuning, the measurements on top physics, on the other hand, are reaching higher precision and providing many intriguing results, which might reveal the mysterious relation between top quarks and Higgs bosons.

%\newpage

\section*{Acknowledgments}

I thank Andreas Bally and Florian Goertz for their collaboration in the early stages of this work. I am also grateful to Hsin-Chia Cheng, Markus Luty, David Marzocca, \'Alvaro Pastor-Guti\'errez, Avik Banerjee, and Gabriele Ferretti for useful discussions. I would like to acknowledge the Mainz Institute for Theoretical Physics (MITP) of the Cluster of Excellence PRISMA+ (Project ID 390831469), for its hospitality and its partial support during the completion of this work.

\bibliography{TOP_Ref}

\end{document}